\documentstyle[12pt]{article}
\advance \textheight by \topskip
\makeatletter
\@addtoreset{equation}{section}
  
\makeatother
\newcommand{\reff}[1]{(\ref{#1})}
\def\reff #1{(\ref{#1})}

\def\ket #1{|{#1}\rangle}

\def\beqn{\begin{eqnarray}}
\def\eeqn{\end{eqnarray}}
\def\beq{\begin{equation}}
\def\eeq{\end{equation}}

\def\non{\nonumber}

\def\Ga{\Gamma}

\def\de{\delta}

\def\({\left(}
\def\){\right)}

\def\+{{\dagger}}

\begin{document}
\title{Supersymmetry of parafermions}
\author{
Sergei Klishevich${}^a$\thanks{E-mail: klishevich@mx.ihep.su},
Mikhail Plyushchay${}^{b,a}$\thanks{E-mail: mplyushc@lauca.usach.cl}\\
{\small ${}^{a}${\it Institute for High Energy Physics, Protvino,
Moscow region, 142284 Russia}}\\
{\small ${}^{b}${\it Departamento de F\'{\i}sica,
Universidad de Santiago de Chile,
Casilla 307, Santiago 2, Chile}}}
\date{}
\maketitle
\vskip-1.0cm
\begin{abstract}
We show that the single-mode parafermionic type systems possess 
supersymmetry, which is based on the symmetry of characteristic 
functions of the parafermions related to the  generalized
deformed oscillator of Daskaloyannis et al. The supersymmetry is
realized in both unbroken and spontaneously broken phases. 
As in the case of parabosonic supersymmetry observed recently 
by one of the authors,
the form of the associated superalgebra depends on the order of 
the parafermion and can be linear or nonlinear in the Hamiltonian.
The list of supersymmetric parafermionic systems includes
usual parafermions, finite-dimensional q-deformed oscillator,
q-deformed parafermionic oscillator and parafermionic oscillator
with internal $Z_2$ structure. 
\vskip2mm
\noindent 
{\it PACS number(s):}  11.10.Kk, 11.10.Lm, 11.30.Pb, 03.65.Pm
\vskip2mm
\noindent
\end{abstract}

\newpage
\section{Introduction}
Recently, it has been established
that the concepts of generalized statistics \cite{kam,poly}
and supersymmetry \cite{susy,wit,cooper},
which can be unified in the form of parasupersymmetry \cite{rs},
turn out to be intimately related. 
This has been done by observation
of hidden supersymmetry in pure parabosonic systems 
with quadratic Hamiltonians \cite{susy_pb}.
This hidden parabosonic supersymmetry 
generally has a structure of the reduced parasupersymmetry.
On the other hand, parasupersymmetric quantum mechanics \cite{rs,cooper}
is formulated in the simplest case in terms
of one bosonic and one parafermionic degrees of freedom.
Therefore, the following natural questions appear:
is it possible to realize supersymmetry 
in pure parafermionic systems and if so, 
what is its general properties
in comparison with parabosonic supersymmetry \cite{susy_pb}? 
Answering these questions constitutes 
the subject of the present paper.

Supersymmetric quantum mechanics \cite{wit,cooper} being a
$(0+1)$-dimensional model of supersymmetric quantum field
theories is formulated in the case of $N=1$ supersymmetry
in terms of one bosonic and one fermionic degrees of freedom. 
In the bosonized version of supersymmetric quantum mechanics
\cite{susybos,gpz}, $N=1$ supersymmetry is realized 
by means of only one bosonic degree of freedom.  
However, the price which is necessary to pay for such minimal realization of
supersymmetry is the essentially nonlinear (nonlocal) structure
of the Hamiltonian and supercharges.
Alternative realization of $N=1$ supersymmetry
is given by the parabosonic Hamiltonians of the simplest normal ordered,
$H_n=a^+ a^-$, or antinormal ordered, $H_a=a^- a^+$, form \cite{susy_pb}.  
In both cases the Hamiltonians are quadratic in parabosonic
operators, but the associated supercharges have essentially
nonlinear structure in $a^\pm$.  The anticommutator of such
supercharges is given by the polynomial of the Hamiltonian,
$\{Q,Q^\+\}={\cal H}(H)$, whose degree is correlated with the order $p$
of parastatistics defined by the relation
$a^-a^+\ket{0}=p\ket{0}$. In the simplest case 
corresponding to $p=2$, the usual linear superalgebra with
$\{Q,Q^\dagger\}=H$ is associated with the single-mode
parabosonic system being in the phase of exact ($H_n$) or
spontaneously broken ($H_a$) supersymmetry.
For revealing the described parabosonic supersymmetries, 
the $R$-deformed Heisenberg algebra \cite{RDHA}
\beq\label{RDHA}
[a^-,a^+]=1+\nu R,\quad \{R,a^\pm\}=0,\quad R^2=1,
\eeq
has been used, which 
at the non-negative integer values of the
deformation parameter, $\nu=p-1$, $p=1,2,\ldots,$ is directly related to a
paraboson of the order $p$ \cite{kam}.  But it was established
\cite{RDHA} that at the special negative values of the deformation
parameter given by the sequence $\nu=-(2k+1)$, $k=1,2,\ldots$, some
parafermionic algebra with internal $Z_2$ structure is naturally
associated with the algebra (\ref{RDHA}).  Therefore, 
this paraboson-parafermion universality of the algebra (\ref{RDHA}) 
indicates on the possibility of
realizing supersymmetry in the single-mode
finite-dimensional parafermionic type systems. 

In Section 2 we consider
finite-dimensional representations of the generalized deformed
oscillator of Daskaloyannis et al \cite{gHA,quesne}
which includes algebra (\ref{RDHA}) as a particular case \cite{susy_pb},
and show that the
systems described by the simplest Hamiltonians $H_n=f^\+f$ and
$H_a=ff^\+$ possess the unbroken supersymmetry,
whereas the Hamiltonian $H_s=f^\+f + ff^\+$
describes supersymmetric systems in the phase of spontaneously broken
supersymmetry.
Here the analysis is given in general form not
depending on the concrete structure of the parafermionic type
oscillator, and it is shown that in the cases of broken and unbroken 
supersymmetries 
the form of the associated superalgebra depends
on the parity of the order $p$ of the (generalized)
parafermion.  
In Section 3 we consider particular examples of
finite-dimensional supersymmetric systems related to various
deformation schemes of the single-mode oscillator algebra.  This
includes the usual parafermions \cite{kam}, the generalized
deformed parafermions with internal $Z_2$ structure \cite{RDHA},
the finite-dimensional q-deformed oscillator
\cite{mac,q-b1,q-b2} and the q-deformed parafermions \cite{q-f1,q-f2}.
We find also that the appropriate linear combination
of the Hamiltonians $H_n$ and $H_a$ in the case of usual
parafermions gives rise to nonlinear supersymmetries
characterized, like in the case of parabosons,
by the presence of arbitrary (fixed)
number of singlet states. Section 4 summarizes the results.

%%%%%%%%%%%%%%%%%%%%%%%%%%%%%%%%%%%%%%%%%%%%%%%%%%%%%%%%%%%%%%%%%%%%%%

\section{Supersymmetry of generalized deformed oscillator}
%%%%%%%%%%%%%%%%%%%%%%%%%%%%%%%%%%%%%%%%%%%%%%%%%%%%%%%%%%%%%%%%%%%%%%

\subsection{Generalized deformed oscillator}

The generalized deformed oscillator \cite{gHA} is defined
by the algebra generated by the operators $\{1,f,f^{\dag},N\}$ 
satisfying the relations
\begin{equation}
\label{gen}
[N,f]=-f,\quad [N,f^{\dag}]=f^{\dag},\quad
f^{\dag}f=F(N),\quad ff^{\dag} = F(N+1),
\end{equation}
where $N$ is the number operator and 
the structure function $F(x)$ is an analytic 
function with the properties
$F(0)=0$ and $F(n)>0$, $n=1,\ldots$. As a consequence, 
the operators $f$, $f^{\dag}$ obey the commutation and
anticommutation relations 
\begin{equation}
\label{com}
\{f^{\dag},f\}=F(N+1)+F(N),\quad 
[f,f^{\dag}]=F(N+1)-F(N).
\end{equation}
The structure function $F(x)$ is characteristic to the deformation
scheme.

The generalized deformed oscillator algebra 
can naturally be represented on the Fock
space of eigenstates of the number operator $N$,
$
N\ket{n}=n\ket{n},\quad \langle n|m\rangle =\de_{nm},
$
$
n=0,1,\ldots,
$
if the vacuum state is defined as $f\ket{0}=0$. 
Such eigenstates are given by 
\[
\ket{n} =\frac1{\sqrt{F(n)!}}(f^\+)^n\ket{0},\quad 
F(n)!\equiv
\prod_{k=1}^nF(k).
\]
The operators $f$ and $f^\dagger$ are 
the annihilation and creation operators of this
oscillator algebra,
\begin{equation}
\label{ancr}
f\ket{n}=\sqrt{F(n)}\ket{n-1},\quad
f^\+\ket{n}=\sqrt{F(n+1)}\ket{n+1}.
\end{equation}
When the characteristic function obeys the condition
\beq
\label{p+1}
F(p+1)=0,
\eeq
the creation-annihilation operators satisfy the relations
$
f^{p+1}=(f^\+)^{p+1}=0.
$
This means that the corresponding 
representation is $(p+1)$-dimensional and 
we have a parafermionic type system of the order $p$.
It is this case that will be treated in what follows.

\subsection{Unbroken supersymmetry}

Let us suppose that the structure function satisfies also 
the property
\beq\label{sym}
F(n)=F(p+1-n).
\eeq
Then the parafermionic-like systems given by 
the quadratic Hamiltonians $H_n=f^\+f$ and $H_a=ff^\+$
possess the degenerated spectra, whose typical form
is shown on \mbox{Fig. 1}.
%%%%%%%%%%%
%%%FIG 1
%%%%%%%%%%%%%%
%%%%%%%%%%%%%%%%%%%%%%

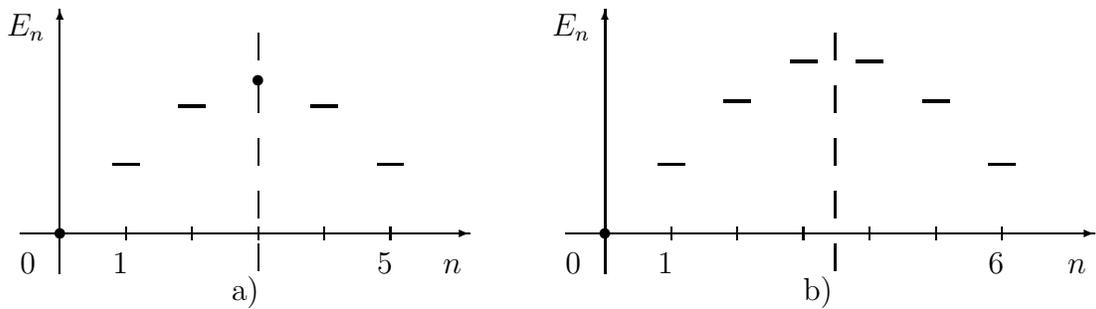
\begin{figure}
\begin{center}
\begin{picture}(170,100)
\put(0,15){\vector(1,0){170}}
\put(15,0){\vector(0,1){100}}
\put(160,0){$n$}
\put(-5,90){$E_n$}
\multiput(15,12.5)(25,0){6}{\line(0,1){5}}
\multiput(90,1)(0,20){5}{\line(0,1){10}}
\put(80,-10){a)}
\put(15,15){\circle*{4}}
\put(35,41){\line(1,0){10}}
\put(35,41.5){\line(1,0){10}}
\put(135,41){\line(1,0){10}}
\put(135,41.5){\line(1,0){10}}
\put(60,63){\line(1,0){10}}
\put(60,63.5){\line(1,0){10}}
\put(110,63){\line(1,0){10}}
\put(110,63.5){\line(1,0){10}}
\put(90,73){\circle*{4}}
\put(0,0){0}\put(35,0){1}\put(135,0){5}
%\multiput(15,41)(10,0){13}{\line(1,0){5}}
%\multiput(15,63)(10,0){10}{\line(1,0){5}}
\end{picture}
\hspace{1cm}
\begin{picture}(200,100)
\put(0,15){\vector(1,0){200}}
\put(15,0){\vector(0,1){100}}
\put(190,0){$n$}
\put(-5,90){$E_n$}
\multiput(15,12.5)(25,0){7}{\line(0,1){5}}
\put(0,0){0}\put(35,0){1}\put(160,0){6}
\multiput(102,1)(0,20){5}{\line(0,1){10}}
\put(90,-10){b)}
\put(15,15){\circle*{4}}
\put(35,41){\line(1,0){10}}
\put(35,41.5){\line(1,0){10}}
\put(60,65){\line(1,0){10}}
\put(60,65.5){\line(1,0){10}}
\put(160,41){\line(1,0){10}}
\put(160,41.5){\line(1,0){10}}
\put(135,65){\line(1,0){10}}
\put(135,65.5){\line(1,0){10}}
\put(85,80){\line(1,0){10}}
\put(85,80.5){\line(1,0){10}}
\put(110,80){\line(1,0){10}}
\put(110,80.5){\line(1,0){10}}
%\multiput(15,41)(10,0){15}{\line(1,0){5}}
%\multiput(15,65)(10,0){13}{\line(1,0){5}}
%\multiput(15,80)(10,0){10}{\line(1,0){5}}
\end{picture}
\vspace{2mm}
\caption{Typical supersymmetric spectra of the parafermionic type
system with the Hamiltonian $H_n=f^\+f$ for the cases of a) odd $p$ ($p=5$) 
and b) even $p$ ($p=6$).}
\end{center}
\label{fig}
\end{figure}
%%%%%%%%%%%%%%%%%

The parafermions \cite{kam}, the
generalized deformed parafermions with internal 
$Z_2$ structure \cite{RDHA}, 
the finite-dimensional q-deformed oscillator
\cite{mac,q-b1,q-b2} and the q-deformed parafermions \cite{q-f1,q-f2}
are given by the characteristic functions
having the property \reff{sym}.
The form of the degenerated spectra indicates that these 
systems may possess the supersymmetry of two types.
The supersymmetry corresponding to the case of even $p$ is characterized 
by the presence of one singlet state 
($\ket{0}$ for $H_n$ and $\ket{p}$ for $H_a$), 
whereas the case of odd $p$ is specified by the presence of 
two singlet states ($\ket{0}$ and $\ket{(p+1)/2}$ for 
$H_n$, and $\ket{p}$ and $\ket{(p-1)/2}$ for $H_a$). 
Therefore, one can expect that different
superalgebras can be associated with
such systems in the cases of the even and of the odd order 
generalized parafermions.

Let us find the corresponding supercharge
operators and the form of superalgebras which they form
with the Hamiltonian.
We start with the case of even $p$.
In the system given by the Hamiltonian
$H_n=f^\+f$, there are $p/2$ pairs of the states
$\{\ket{n},\ \ket{p+1-n}\}$ 
with energy $E_n=E_{p+1-n}=F(n)$, $n=1,\ldots,p/2$. 
The supercharges $Q$ and $Q^\+$ have to transform mutually
the states in each supersymmetric doublet and annihilate the 
singlet state $\ket{0}$. One can
specify them by the relations
\begin{eqnarray}
&Q\ket{0}=0,\quad Q\ket{p+1-n}\sim\ket{n},\quad
Q\ket{n}=0,&\nonumber\\
&Q^\+\ket{0}=0,\quad 
Q^\+\ket{n}\sim\ket{p+1-n},\quad
Q^\+\ket{p+1-n}=0,&
\label{Qact}
\end{eqnarray}
where $n=1,...,p/2$. These relations mean that
the  operator $Q^\+$ transforms
the first half of the states forming supersymmetric doublets
into the second half states while the conjugate operator $Q$
acts in the opposite direction. 
Eqs.  (\ref{Qact}) and (\ref{sym}) lead to the relations
\beq\label{q^2}
Q^2=(Q^\+)^2=0,\quad [Q,H_n]=[Q^\+,H_n]=0,
\eeq
meaning that the operators $Q$ and $Q^\+$ 
are the mutually conjugate conserving 
nilpotent operators, i.e. are the supercharges.
In correspondence with Eqs. (\ref{Qact}), (\ref{q^2}),
they can be represented in the form
\beq\label{Qanz}
Q=\sum_{n=1}^{p/2}C_n^pP_n^pf^{p+1-2n},
\eeq
where $P_n^p$ are the projectors,
$P_n^p|m\rangle =\delta_{nm}|n\rangle,$
$P_n^pP_m^p=\delta_{nm}P_n^p,$
$\sum_{n=0}^pP_n^p=1,$
and the coefficients $C_n^p$ have to be specified.
The projectors
can be realized in terms of the number operator,
\begin{equation}\label{proj}
P_n^p = \frac{(-1)^{p+n}}{n!(p-n)!}\mathop{{\prod}'}_{k=0}^p(N-k),
\quad n=0,...,p,
\end{equation}
where the prime means that the term with $k= n$ is omitted.
Using the identities
\[
(f^\+)^nf^n =\prod_{m=0}^{n-1}F(N-m),\quad
f^n(f^\+)^n =\prod_{m=1}^{n}F(N+m),\quad
\prod_{n=0}^p(N-n)=0,
\]
one can calculate the anticommutator of the operators $Q$ and $Q^\+$:
\begin{eqnarray}\label{acom}
\non
\{Q,Q^\+\}&=&
\sum_{n=1}^{p/2}\(P_n^p+P_{p+1-n}^p\)\(C_n^p\)^2\prod_{m=0}^{p-2n}F(n+
m)
\\&=&
F(N)\sum_{n=1}^{p/2}\(P_n^p+P_{p+1-n}^p\)
\(C_n^p\)^2\prod_{m=1}^{p-2n}F(n+m).
\end{eqnarray}
Here we have used the relation
\beq\label{prop}
g(n)P_n^p=g(N)P_n^p,
\eeq
taking place for arbitrary function $g(N)$. 
{}From Eq. \reff{acom} one can find that
the choice
\beq\label{fact}
C_n^p=\frac{F(n)!}{F(\frac{p}{2})!},\quad n=1,\ldots,
\frac{p}{2},
\eeq
leads to the usual $N=1$ superalgebra given by Eq. (\ref{q^2}) and
\beq
\label{n1}
\{Q,Q^\+\}=H_n.
\eeq

The supercharges given by Eqs. (\ref{Qanz}) and (\ref{fact})
can be represented in a more compact form. 
For the purpose, let us introduce the $Z_2$-grading operator $\Ga$
defined by the relations
$\Ga^2=1$, $[\Ga,H]=0$, $\{\Ga,Q\}=\{\Ga,Q^\dagger\}=0$.
It can be chosen in the form
\beq\label{gr_op}
\Ga=\left(
\sum_{n=0}^{\frac{p}{2}}-\sum_{n=\frac{p}{2}+1}^{p}\right)
P_n^p.
\eeq
Then the  states $\ket{n}$ with $n=0,\ldots,\frac{p}{2}$
and $n=\frac{p}{2},\ldots,p$ 
have the parities $+1$ and $-1$, respectively. 
On the other hand, 
the operators $f$ and $f^\+$ do not have definite parity
(they are neither even nor odd operators). 
Using the relation \reff{prop},
one can represent the supercharges in the form
\beq\label{QP}
Q=C^p(N)f^{2N-p-1}P_-,
\quad
Q^\+ =(f^\+)^{p+1-2N}C^p(N)P_+,
\eeq
where $P_\pm=\frac12\(1\pm\Ga\)$ are the projectors on the 
even and odd parity subspaces,
and $C^p(N)$ is given by Eq. (\ref{fact}) with 
$F(n)$ replaced by $F(N)$.

By introducing the operators
\beq
\label{fftil}
\tilde{f}=(F(N+1))^{-1/2}f,\quad
\tilde{f}{}^\dagger=(F(N))^{-1/2}f^\dagger,
\eeq
satisfying the relations 
$
\tilde{f}{}^{p+1}=(\tilde{f}{}^\dagger)^{p+1}=0,
$
$
(\tilde{f}{}^\dagger)^{n}\tilde{f}{}^n=\sum_{k=n}^pP_k^p,
$
$
\tilde{f}{}^{n}(\tilde{f}{}^\dagger)^n=\sum_{k=0}^{p-n}P_k^p,
$
one can also represent the supercharges as
\beq\label{ftil}
Q=\sqrt{\varphi (N)}\tilde{f}{}^{2N-p-1}P_-,\quad
Q^\dagger=\sqrt{\varphi (N)}(\tilde{f}{}^\dagger)^{p+1-2N}P_+,
\eeq
with $\varphi(N)=F(N)$.

Now let us turn to the case of odd $p$,
where the system described by the Hamiltonian $H_n=f^\dagger f=F(N)$
is characterized by the presence of two singlet (unpaired) 
states $\ket{0}$ and $\ket{\frac{p+1}{2}}$ with energies $E_0=0$ and 
$E_{\frac{p+1}{2}}=F(\frac{p+1}{2})$.
The operator
\beq\label{Qanz1}
Q=\sum_{n=1}^{\frac{p-1}{2}}C_n^pP_n^pf^{p+1-2n},
\eeq
and its conjugate, $Q^\dagger$,
annihilate both singlet states,
satisfy relations (\ref{Qact}) with 
$n=1, \ldots,\frac{p-1}{2}$,
and obey Eq. (\ref{q^2}). 
To fix the unknown coefficients $C^p_n$,
we calculate the anticommutator of the operators $Q$ and $Q^\dagger$:
\begin{eqnarray}
\label{acom1}
\non
\{Q,Q^\+\}&=&
F(N)\sum_{n=1}^{\frac{p-1}{2}}\(P_n^p+P_{p+1-n}^p\)
\(C_n^p\)^2\prod_{m=1}^{p-2n}F(n+m).
\end{eqnarray}
Therefore, assuming that $F(\frac{p+1}{2})>F(n)$,
$n\neq \frac{p+1}{2}$, and imposing the condition
\beq\label{al}
\(C_n^p\)^2\prod_{m=1}^{p-2n}F(n+m)=\Delta (n),
\quad
\Delta (n)\equiv
 F({\textstyle\frac{p+1}2}) - F(n),
\eeq
we get the relation
\beq\label{alg_n}
\{Q,Q^\+\} = H_n\(E_{\frac{p+1}2} - H_n\).
\eeq
Taking into account the symmetry relation (\ref{sym}), 
we get the solution to Eq. (\ref{al}):
\beq\label{c_n}
C_n^p =\sqrt{\Delta(n)/E_{\frac{p+1}2}}\cdot
\frac{F(n)!}{F(\frac{p-1}{2})!},
\quad n=1,\ldots,{\textstyle\frac{p-1}{2}}.
\eeq
The $Z_2$-grading operator commuting with the Hamiltonian
and anticommuting with the supercharges can be chosen here in the form
\beq\label{gr_op2}
\Ga=\left(\sum_{n=0}^{\frac{p-1}{2}}-\sum_{n=\frac{p+1}{2}}^{p}\right)
P_n^p.
\eeq
With its help we represent the supercharges (\ref{Qanz1}) in
the form \reff{QP}, where now the projectors $P_\pm$
are defined in terms of grading operator (\ref{gr_op2}),
and $C^p(N)$ is given by Eq. (\ref{c_n})
with $F(n)$ replaced by $F(N)$.
In terms of the operators (\ref{fftil})
the supercharges can also be given in the form 
(\ref{ftil}) with $\varphi(N)=F(N)\Delta(N)$,
where $\Delta(N)$ is given by Eq. (\ref{al})
with $F(n)$ substituted for $F(N)$.

Therefore, for odd $p$ the system $H_n=f^\dagger f$
can be characterized by the nonlinear superalgebra (\ref{q^2}), (\ref{alg_n}),
and we conclude that in general case of the even and odd order generalized 
parafermions the anticommutator of supercharges
is given by the polynomial of the Hamiltonian
whose order coincides with the number of singlet states.
This resembles the property of the supersymmetry in pure parabosonic systems
with normal ordered Hamiltonian where the order of
the polynomial appearing in the anticommutator of supercharges
coincides with the number of singlet states in the system,
which, in turn, is defined by the order of the paraboson.

However, here, unlike the parabosonic case,
the systems given by $H_n$ and $H_a$ reveal,
in fact, the same supersymmetry.
Indeed, due to the symmetry relation (\ref{sym}),
the spectrum of the Hamiltonian $H_a=ff^\+=
F(N+1)$ is a mirror reflection of the spectrum 
of the Hamiltonian $H_n=f^\+f=F(N)$ with respect to the middle point
of the interval $[0,p]$.
As a consequence, all the analysis and formulas for 
the supersymmetric parafermionic systems described by $H_a$ 
may be immediately obtained from those given above
for the systems with $H_n$
via the formal substitutions
$f\rightarrow f^{\dag },$ 
$f^{\dag }\rightarrow f,$
$n
\rightarrow p-n,
$
and we do not expose them explicitly.

%%%%%%%%%%%%%%%%%%%
\subsection{Spontaneously broken supersymmetry}
%%%%%%%%%%%%%%%%%%%

Let us consider the parafermionic system 
given by the quadratic Hamiltonian
\begin{equation}
\label{Hs}
H_s=f^\+f + ff^\+.
\end{equation}
When the associated characteristic function
satisfies the symmetry relation (\ref{sym}),
the spectrum of (\ref{Hs}) is also supersymmetric
(see Fig. 2). However, this time the spectrum has no
energy zero level, but, instead, the lowest 
energy level is positive and degenerated: $E_0=E_p=F(1)>0$.
Therefore, the system described by the Hamiltonian
(\ref{Hs}) has the features of the system realizing
spontaneously broken supersymmetry.
Here, as for the supersymmetric systems given by $H_n$ and $H_a$, 
two cases have to be distinguished:
with odd $p$'s all the states are paired in supersymmetric
doublets, whereas the case of even $p$'s is characterized 
by the presence of one singlet state $\ket{\frac{p}{2}}$. 
%%%%%%%%%%%%%
%%%FIG 2
%%%%%%%%%%%%%
%%%%%%%%%%%%%%%%%%%%%%%%%%%%%%%%
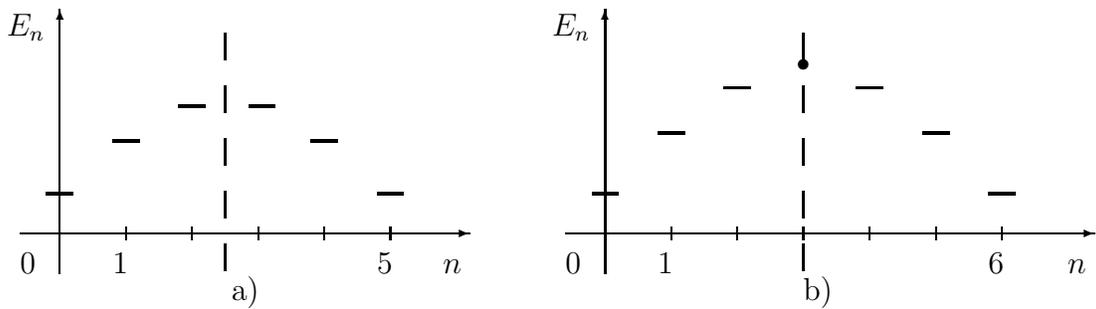
\begin{figure}
\begin{center}
\begin{picture}(170,100)
\put(0,15){\vector(1,0){170}}
\put(15,0){\vector(0,1){100}}
\put(160,0){$n$}
\put(-5,90){$E_n$}
\multiput(15,12.5)(25,0){6}{\line(0,1){5}}
\multiput(77.5,1)(0,20){5}{\line(0,1){10}}
\put(80,-10){a)}
\put(10,30){\line(1,0){10}}
\put(10,30.5){\line(1,0){10}}
\put(135,30){\line(1,0){10}}
\put(135,30.5){\line(1,0){10}}
\put(35,50){\line(1,0){10}}
\put(35,50.5){\line(1,0){10}}
\put(110,50){\line(1,0){10}}
\put(110,50.5){\line(1,0){10}}
\put(60,63){\line(1,0){10}}
\put(60,63.5){\line(1,0){10}}
\put(86.5,63){\line(1,0){10}}
\put(86.5,63.5){\line(1,0){10}}
\put(0,0){0}\put(35,0){1}\put(135,0){5}
\end{picture}
\hspace{1cm}
\begin{picture}(200,100)
\put(0,15){\vector(1,0){200}}
\put(15,0){\vector(0,1){100}}
\put(190,0){$n$}
\put(-5,90){$E_n$}
\multiput(15,12.5)(25,0){7}{\line(0,1){5}}
\put(0,0){0}\put(35,0){1}\put(160,0){6}
\multiput(90,1)(0,20){5}{\line(0,1){10}}
\put(90,-10){b)}
\put(10,30){\line(1,0){10}}
\put(10,30.5){\line(1,0){10}}
\put(160,30){\line(1,0){10}}
\put(160,30.5){\line(1,0){10}}
\put(35,53){\line(1,0){10}}
\put(35,53.5){\line(1,0){10}}
\put(135,53){\line(1,0){10}}
\put(135,53.5){\line(1,0){10}}
\put(60,70){\line(1,0){10}}
\put(60,70.5){\line(1,0){10}}
\put(110,70){\line(1,0){10}}
\put(110,70.5){\line(1,0){10}}
\put(90,79){\circle*{4}}
\end{picture}
\vspace{2mm}
\caption{Typical spectra of the parafermionic system with the Hamiltonian
$H_a=f^\+f + ff^\+$ in the cases of 
a) odd $p$ ($p=5$) and b) even $p$ ($p=6$).}
\end{center}
\end{figure}
%%%%%%%%%%%%%%%%%%%%%%%%%

The supercharge for the system (\ref{Hs}) can be realized in the form
$$
Q=\sum\limits_{n=0}^{[\frac{p+1}{2}]-1}C_n^pf^{p-2n}P_{p-n}^p, 
$$
where $[.]$ means the integer part and
the coefficients $C_n^p$ are defined by
\beq\label{cpbr}
C_n^p=\sqrt{\chi(n)}\cdot \frac{F(n)!}{F([\frac{p+1}{2}]-1)!},
\quad
n=0,\ldots,{\textstyle[\frac{p+1}{2}]-1},
\eeq
with $\chi(n)=E_n/F(\frac{p+1}{2})$ for odd $p$ and
$\chi(n)=E_{\frac{p}{2}}-E_n$ for even $p$;
$E_n$ is the energy of the state with number $n$  and 
we suppose that for even $p$, $E_\frac p2$ is the maximum energy.

The supercharge $Q$ and its conjugate obey the relations
\beq
\label{bro}
\{Q,Q^\+\}=H_s \quad \mbox{(odd }p),\quad 
\{Q,Q^\+\}=E_{\frac p2}-H_s \quad \mbox{(even }p).
\eeq
Introducing the grading operator
\beq
\label{rbr}
\Gamma =\left(\sum\limits_{n=0}^{[\frac{p}2]}-
\sum\limits_{n=[\frac{p}{2}]+1}^{p}\right)
P_n^p, 
\eeq
one can represent the supercharges in the form
\[
Q=C^p(N)f^{2N-p}P_-,\quad
Q^\+=(f^\+)^{p-2N}C^p(N)P_+,
\]
where $C^p(N)$ is obtained from Eq. (\ref{cpbr})
with $n$ substituted for $N$.
The operators (\ref{fftil}) give the alternative
representation
$$
Q=\sqrt{\varphi(N)}\tilde{f}{}^{2N-p}P_{-},\quad
Q^\+=\sqrt{\varphi(N)}(\tilde{f}{}^\+)^{p-2N}P_+,
$$
where $\varphi(N)=F(N)+F(N+1)=H_s$ for odd $p$
and $\varphi(N)=E_{\frac{p}{2}}-H_s$ for even $p$.

Thus, here in both cases of even and odd $p$'s the anticommutator
of supercharges is linear in Hamiltonian $H_s$,
but its form is different due to 
different structure of the corresponding spectra.

%%%%%%%%%%%%%%%%%%%%%%%%%%%%%%%%%%%%%%%%%%%%%%%%%%%%%%%%%%%%%%%%%%%%%%
\section{Finite-dimensional supersymmetric systems}
%%%%%%%%%%%%%%%%%%%%%%%%%%%%%%%%%%%%%%%%%%%%%%%%%%%%%%%%%%%%%%%%%%%%%%
In this section we consider the particular examples of supersymmetric 
finite-dimensional systems related to various deformation schemes 
of the oscillator algebra, for which the results of the previous section
can be applied.

\subsection{Parafermionic oscillator}
The algebra of the usual parafermions \cite{kam} is defined by the
relations
\[
\{f,f^\+\}=p + 2pN - 2N^2,\quad [f,f^\+]=p-2N,
\]
from which the trilinear parafermionic relations,
$[[f^\+,f],f^\+]=2f^\+$, 
$[[f^\+,f],f]=-2f$,
and the vacuum relation $ff^\+\ket{0}=p\ket{0}$ can be immediately obtained.
The comparison with Eq. (\ref{com})
gives the parafermionic structure function:
\beq\label{0par}
F(N)=N(p+1-N).
\eeq
It possesses the symmetry
property \reff{sym} which leads to the discussed supersymmetries. 

\subsection{Finite-dimensional q-deformed oscillator}
The q-deformed oscillator is defined by the relation
\cite{mac,q-b1,q-b2}
$$
aa^\+-q^{-1}a^\+a=q^N.
$$
The structure function of this system is 
\beq
\label{qos}
F(N)=F_q(N)= \frac{q^N-q^{-N}}{q-q^{-1}}.
\eeq
When $q$ is a primitive root of unity of the even order,
$q^{2(p+1)}=1$, i.e.
$q=e^{i\frac{\pi}{p+1}}$,
the system has  the $(p+1)$-dimensional Fock type representation, 
characterized by the relations
$
a^{p+1}=(a^\+)^{p+1}=0.
$
In this case the structure function acquires the 
form
\beq\label{sin}
F(N)=\frac{\sin\frac{\pi N}{p+1}}{\sin\frac{\pi}{p+1}},
\eeq
and obeys the symmetry condition \reff{sym}.

\subsection{The q-deformed parafermionic oscillator}
The q-deformed parafermionic algebra \cite{q-f1,q-f2} is defined by
the structure function
$$
F(N)=F_q(N)F_q(p+1-N),
$$
where $F_q(N)$ is given by Eq. (\ref{qos}).
This structure function obeys the symmetry relation
(\ref{sym}).  
At $q=e^{i\frac{\pi}{p+1}}$ we have the $(p+1)$-dimensional representation
with the structure function which, due to Eq. (\ref{sym}),
is reduced to the square of the structure function (\ref{sin}):
\beq\label{sin2}
F(N) = \frac{\sin^2\frac{\pi N}{p+1}}{\sin^2\frac{\pi}{p+1}}.
\eeq

All three systems (\ref{0par}), (\ref{sin}) and (\ref{sin2})
possess the spectra corresponding to the exact (unbroken)
and spontaneously broken supersymmetry of the type shown on  Figs. 1
and 2. Note that in this context 
the fermion system ($p=1$, $F(N)=N(2-N)$) with Hamiltonian
$H_s=\{f^\+,f\}=1$ realizes the simplest case of 
spontaneously broken supersymmetry.

\subsection{Parafermionic oscillator with internal $Z_2$ structure} 

The even order parafermionic oscillator with
internal $Z_2$ structure is associated with 
the $R$-deformed Heisenberg algebra which,
in turn, is related to parabosons \cite{RDHA}.
The $R$-deformed Heisenberg algebra is generated by the operators
$\{1,a^\pm,R\}$ satisfying the (anti)commutation relations
(\ref{RDHA}),
where $\nu\in {\bf R}$ is a deformation parameter and $R$ is the
reflection operator. The structure function of this algebra is
\beq\label{rd}
F(N)=N+\nu\frac{1}{2}(1-(-1)^N)=N+\nu\sin^2\frac{\pi N}{2}.
\eeq
On the half-line $\nu>-1$, it is positive definite,
$F(n)>0$ for $n=1,2,\ldots$,
that means that the algebra \reff{RDHA} 
with $\nu>-1$ has infinite-dimensional unitary
representations. 
When $\nu=p-1$, $p=1,2,\ldots$,
they are the representations of the parabosonic 
oscillator system of the order $p$. 
In the case of odd $\nu$'s, $\nu=2k+1$, $k=0,\ldots$,
the structure function (\ref{rd}) satisfies the 
relation 
$
F(2n+1)=F(2n+\nu+1)
$
underlying the supersymmetries 
of the pure parabosonic systems given by the Hamiltonians
$H_n=a^+a^-=F(N)$ and $H_a=a^-a^+=F(N+1)$ \cite{susy_pb}.

Though the algebra \reff{RDHA} emerges from generalizing the
bosonic harmonic oscillator system \cite{kam}, it has also the 
finite-dimensional representations of the 
parafermionic nature \cite{RDHA}. 
Indeed, at the special
values of the deformations parameter, $\nu={} - (p+1),\ p=2,4,6,...$
the structure function (\ref{rd}) satisfies the
relation $F(p+1)=0$, which signals on existence of
$(p+1)$-dimensional irreducible representations of the algebra
\reff{RDHA}, in which the relations $(a^\pm)^{p+1}=0$ are valid. 
However, in this case 
the structure function $F(N)$ is not positive-definite,
and the corresponding finite-dimensional
representations are not unitary. In such representations
the creation-annihilation operators $a^\pm$ 
are mutually conjugate with respect to the
indefinite scalar product. On the other hand, with respect
to the usual positive definite scalar product they are related as
$(a^+)^\dagger=a^-R$ \cite{RDHA}. Therefore, defining new creation-annihilation
operators $f=a^-R$, $f^\+=a^+$, we arrive at the 
following parafermionic type algebra of the even order:
\beq\label{ffR}
\{f,f^\+\}=p+1-R,\quad f^{p+1}=(f^\+)^{p+1}=0,\quad
\{R,f\}=\{R,f^\+\}=0,\quad p=2k.
\eeq
Due to the presence of the reflection operator $R=(-1)^N$,
the algebra has the natural internal $Z_2$ structure.
The structure function of this parafermionic type system is
\beq\label{z2}
F(N)=N(-1)^N+(p+1)\sin^2\frac{\pi N}2.
\eeq
The property \reff{sym} is valid for the function (\ref{z2}). 
Since this finite-dimensional  oscillator with 
internal $Z_2$ structure has the parafermionic 
type algebra of the even order, the corresponding unbroken
supersymmetry (for $H_n$ and $H_a$) 
is characterized by the presence of only one singlet state.
In the simplest case $p=2$ the spectrum
$E_n=(0,2,2)$ of $H_n$ has a form of the type shown on Fig. 1b.
For $p=2k$, $k>1$, the spectrum of $H_n$
has a non-typical $X$-like form,
being a superposition of the increasing and decreasing spectra
of the subspaces of even and odd states,
$E_{2l}=2l$, $l=0,...,k$, and
$E_{2l+1}=2(k-l)$, $l=0,...,k-1$.

On the other hand, due to Eq. (\ref{ffR}), here
the Hamiltonian $H_s=\{f^+,f\}$ has 
a two-level degenerated spectrum for all $p=2k$.
Therefore, 
the present parafermionic system 
has a spectrum of spontaneously broken $N=1$ supersymmetry
only in the simplest case $p=2$,
whereas the cases $p=4,6,\ldots,$ correspond to 
spontaneously broken extended supersymmetry.

\subsection{Generalized supersymmetry with several singlet states}
As we have seen in the previous subsection,
in the case of the Hamiltonian
$H_s$ the parafermionic oscillator system 
with internal $Z_2$ structure is different
from other systems: here only $p=2$ corresponds to
spontaneously broken $N=1$ supersymmetry,
while the
cases $p=4,6,\ldots$ 
give spontaneously broken extended supersymmetry.
The specific nature of the structure function
$F(N)$ can reveal itself in a particular way
in other cases too. 
Let us show that  
with the usual parafermions (\ref{0par})
one can realize the specific supersymmetries
generalizing those already discussed.
They are characterized by the presence of arbitrary (but fixed)
number of singlet states which defines the nonlinear order 
of the corresponding superalgebra.

Such generalized supersymmetries can be found 
in the systems given by the Hamiltonian
\beq
\label{hk}
H_k=\{f^\+,f\} +k[f^\+,f]+p(\vert k\vert-1),\quad
k\in Z,\quad p\geq \vert k\vert +1.
\eeq
At $k=-1,0,1,$ the Hamiltonian $H_k$ is reduced to
$2H_a$, $H_s-p$ and $2H_n$, respectively,
and for $k\neq 0$ it has the spectra
which contain $\vert k\vert$ ($\vert k\vert -1$)
positive energy singlet states in addition 
to zero energy ground state
when $p-k$ is even (odd). 
We note here that analogous
generalized supersymmetric Hamiltonian 
can also be constructed for the case 
of finite-dimensional q-deformed oscillator
(\ref{qos}).
For $k>0$ the corresponding supercharges associated 
with (\ref{hk}) can be 
realized in the form similar to (\ref{ftil}),
\[
Q=\sqrt{\varphi(N)}\tilde{f}{}^{2N-p-k}P_-,\quad
Q^\+=\sqrt{\varphi(N)}(\tilde{f}{})^{p+k-2N}P_+,
\]
with
\[
\varphi(N)={\cal H}\equiv
h(H_k)\prod_{m=0}^{\vert k\vert-1}
(H_k-E_m),
\]
where $E_m$ is the energy of the state $\ket{m}$,
$h(H_k)=1$ when $p-k$ is odd, $h(H_k)=E_{max}-H_k$
for even $p-k$ and $E_{max}=E_{\frac{p+k}{2}}$.
The anticommutator of supercharges is given here by
\[
\{Q,Q^\+\}={\cal H}, 
\]
i.e. it is a polynomial
in $H_k$, whose order coincides with the number of singlet
states. The case $k<0$ is obtained via the 
formal change $f^\+\rightarrow f$, $f\rightarrow f^\+$,
$m\rightarrow p-m$.

%%%%%%%%%%%%%%%%%%%%%%%%%%%%%%%%%
\section{Summary and outlook}
We have shown that the single-mode parafermionic type systems
possess supersymmetry.
It is based on the symmetry relation $F(n)=F(p+1-n)$
of the characteristic function of the known
generalized parafermions related to the deformed oscillator 
(\ref{gen})
of Daskaloyannis et al  \cite{gHA,quesne}.
The simplest quadratic Hamiltonians $H_n=f^\+f=F(N)$ and 
$H_a=ff^\+=F(N+1)$ describe the systems with unbroken
supersymmetry specified by the presence of 
one zero energy ground state in the spectrum,
while the Hamiltonian $H_s=f^\+f + ff^\+$ represents
the system in the phase of spontaneously broken supersymmetry,
characterized by the doublet of positive energy ground states.
Besides, the form of supersymmetry depends 
on whether the order $p$ of the parafermion
is even or odd. In the case of the systems 
with exact supersymmetry for even $p$ there is only  
one singlet state (of zero energy) in the system's spectrum 
($\ket{0}$
for $H_n$ and $\ket{p}$ for $H_a$)
and the anticommutator of the supercharges is equal
to the Hamiltonian, i.e. we have here the usual superalgebra
(\ref{q^2}), (\ref{n1}) 
of the $N=1$ supersymmetry.
In the case of odd $p$, the systems with unbroken supersymmetry
possess another singlet state with positive energy, 
$\ket{\frac{p+1}{2}}$ for $H_n$, or  $\ket{\frac{p-1}{2}}$ for $H_a$,
in addition to the zero energy ground state.
As a consequence, for odd $p$'s the anticommutator of the supercharges
is given by the polynomial quadratic in the Hamiltonian 
(see Eq. (\ref{alg_n})).
The supersymmetries of the systems given by
the normal ordered Hamiltonian $H_n$ and by the antinormal
ordered Hamiltonian $H_a$ turn out
to be the same due to the relation $F(n)=F(p+1-n)$
underlying supersymmetry.
In the case of spontaneously broken
supersymmetry, the case of odd $p$ is characterized
by the presence of only supersymmetric doublets in the spectrum,
whereas the spectra of parafermionic systems of even order $p$ 
contain one singlet state $\ket{\frac{p}{2}}$ 
with positive energy. In both cases the 
superalgebra is linear in $H_s$, but its concrete form
is different (see Eq. (\ref{bro})).
With the quadratic Hamiltonian (\ref{hk}),
one can generalize the parafermionic supersymmetries of the systems 
with $H_n$, $H_a$: in this case 
the spectra are characterized
by the presence of $|k|$ or $|k|-1$ singlet states in addition 
to zero energy ground state and
the superalgebra is characterized by the corresponding polynomial
in $H_k$ appearing in the anticommutator of supercharges $Q$ and $Q^\dagger$.
These properties are similar to those characterizing
supersymmetric pure parabosonic systems \cite{susy_pb}.
As in the case of parabosonic supersymmetry, 
a peculiar nature of the described supersymmetries
of parafermionic systems with quadratic Hamiltonians
is encoded in nonlinear structure
of the corresponding supercharges being represented in 
terms of only creation-annihilation operators $f^\+$, $f$.

It is interesting to note here that similar 
nonlinear superalgebraic structure with the
anticommutator of the supercharges being
quadratic in the Hamiltonian was observed earlier by Isaev and Malik 
in the system of $q$-deformed bosonic and fermionic oscillators \cite{isaev}.
Like in the present case, their
Hamiltonian is quadratic in creation-annihilation
operators and looks like the Hamiltonian of uncoupled
(deformed) bosonic and fermionic oscillators.
Moreover, like in the usual superoscillator (but unlike
the systems treated here),
the odd supercharges of the supersymmetric system of ref. \cite{isaev}
are quadratic in bosonic-fermionic operators,
but the nonlinearity turns out to be encoded in a
nontrivial mutual coupling of the oscillators via
the corresponding  q-deformed commutation relations.

The observed supersymmetries of parafermionic
systems can be used for construction of new types
of parasupersymmetry \cite{Bec}. Indeed, in the simplest case
of the Hamiltonian $H_{ps}=b^\+ b +H$,
where $b^\+$, $b$ are the bosonic creation-annihilation 
operators and $H$ is the Hamiltonian $H_n$, $H_a$ or $H_s$ 
of the usual parafermions (\ref{0par}),
we get the system with parasupersymmetric spectrum.
In the lower part the spectrum is different
from that of the 
Rubakov-Spiridonov parasupersymmetric system
$H_{psusy}=b^\+ b+\frac{1}{2}([f^\+,f]+p)$ \cite{rs}.
For the same purpose,
one can use the appropriate linear combination 
of one of the supersymmetric parabosonic Hamiltonians 
with one of the supersymmetric parafermionic Hamiltonians.

We have treated the parafermionic
systems given by the Hamiltonians of the 
simplest quadratic form.
However, it is clear that the Hamiltonians
of the form ${\cal H}={\cal F}(H)$
with arbitrary function ${\cal F}$ of
$H=H_n,H_a,H_s$
also possess the supersymmetric spectra.
The same is true for the analogous generalization
of the parabosonic systems.
Therefore, the observed supersymmetries of 
the single-mode parabosonic and parafermionic
systems survive under generalization
to the case of the `self-interacting' systems.
It seems that generalizing the present single-mode
constructions to the many-modes oscillator systems,
one could arrive at the physically interesting 
supersymmetric quantum field systems, 
formulated, e.g., in terms of only one parabosonic
or parafermionic field.

\vskip0.5cm
{\bf Acknowledgements}
\vskip3mm

The work of M.P. was supported in part by 
the grant 1980619 from FONDECYT (Chile)
and by DICYT (USACH).

%%%%%%%%%%%%%%%%%%%%%%%%%%%%%%%%%

\end{document}